# Theoretical investigation of thermodynamic balance between cluster isomers and statistical model for predicting isomerization rate

Zheng-Zhe Lin*

*School of Physics and Optoelectronic Engineering, Xidian University, Xi'an 710071, China*

*Corresponding Author. E-mail address: linzhengzhe@hotmail.com

**Abstract** – By molecular dynamics simulations and free energy calculations based on Monte Carlo method, the detailed balance between Pt cluster isomers was investigated. For clusters of $n \leq 50$, stationary equilibrium is achieved in 100 ns in the canonical ensemble, while longer time is needed for $n > 50$. Then, a statistical mechanical model was built to evaluate unimolecular isomerization rate and simplify the prediction of isomer formation probability. This model is simpler than transition state theory and can be easily applied on *ab initio* calculations to predict the lifetime of nanostructures.

## I. Introduction

In modern nanotechnology, thermal stability of specific atomic geometry configurations is essential to the design of nanostructures. Before preparing a nanostructure, theoretical prediction of the stability should be essential to avoid repeated experimental attempts. Since a long time ago, to solve problems about clean energy conversion and storage, great efforts have been focused on finding stable Pt-based nano-catalyst. In recent years, bimetallic Pt$_3$M [1, 2] and multimetallic Au/FePt$_3$ nanoparticles [3] with high electrocatalytic activity were developed. And Pt, Rh and Pd have been used very extensively in heterogeneous catalysis, especially for reactions involving CO and H$_2$. To predict stable configuration of Pt nanoparticles, theoretical simulations have been focused on the potential energy and thermal evolution [4-6]. In theoretical point of view, the growth and isomerization of nanoparticles belong to thermal-driven atomic migrations. Therefore, corresponding theoretical investigations should emphasize on the kinetics of formation and isomerization reactions.

Predicting the shape of nanoparticles is not a simple task because their growth involves many atomic processes. In equilibrium, the cluster or nanoparticle isomer with higher formation probability corresponds to lower free energy. For isothermal-isobaric situation, at temperature $T$ the chemical balance between isomer $a$ and $b$ satisfies

$$G_b - G_a = -kT\ln(N_b/N_a), \tag{1}$$

where $G$ the Gibbs free energy of one molecule and $N$ the molecule number [7], as well as for isothermal-isovolumic situation the Helmholtz free energy $F$ satisfies

$$F_b - F_a = -kT\ln(N_b/N_a). \tag{2}$$

However, when $T$ is not high enough, the transformation rates between isomers are too slow for the system to reach equilibrium, and the free energy criterion is no longer tenable. For example, for three-dimensional crystals or two-dimensional islands on solid surfaces formed away from equilibrium, their shapes are away from the prediction of free-energy-based Wulff construction [8-14]. It has been found that the evolution of Co islands is thermodynamically dominated at 300~600 K, while Pt and Pd islands are kinetically dominated at the same temperature [15].

Either in equilibrium or non-equilibrium, the evolution of isomer number $N$ can be evaluated by solving kinetics equations based on unimolecular isomerization rates. In equilibrium $N$ becomes time-independent and satisfies the detailed balance principle [16]. At high molecular concentration, unimolecular reactions are dominated by intermolecular collisions and present $1^{st}$-order behavior, and in equilibrium the detailed balance principle reads

$$N_b/N_a = k_{1st\ a\to b}/k_{1st\ b\to a}, \tag{3}$$

where $k_{1st}$ is isomerization rate constant. Theoretical prediction of $k_{1st}$ should be an efficient way to calculate isomer formation probability because it may be simpler than free energy calculations and can be applied on non-equilibrium case.

In this work, isomerization of Pt clusters with dozens of atoms was investigated by molecular dynamics (MD) simulations, and stationary equilibrium and detailed balance in the thermodynamic evolution were verified. In the MD, the number of each isomer in the canonical ensemble is in good agreement with theoretical value (Eq. (1) or (2)) calculated by free energy obtained from a technique combining rigid-rotor and harmonic-oscillator approximation and Monte Carlo method. Then, a statistical mechanical model was built to evaluate unimolecular isomerization rate, and its accuracy was validated. By the results, our model produces similar rate with transition state theory (TST), and this model is simple to be applied on *ab initio* calculations.

**II. MD simulations**

To collect data for investigating cluster isomerization, a model of vapor-phase Pt cluster growth was set up by MD simulation. A tight-binding like potential $U = \sum_{i=1}^{n} U_i$ for interaction between Pt atoms was employed, in which the energy of $i^{\text{th}}$ atom in a $n$-atom system is written as

$$U_i = \sum_{j=1, j \neq i}^{n} A e^{-p(r_{ij}/r_0 - 1)} - \left( \sum_{j=1, j \neq i}^{n} \xi^2 e^{-2q(r_{ij}/r_0 - 1)} \right)^{1/2}, \quad (4)$$

where $r_{ij}$ denotes the distance between the $i^{\text{th}}$ and $j^{\text{th}}$ atom and $A$, $p$, $q$, $\xi$ and $r_0$ are presented in Ref. [17]. He atoms were used as buffer gas, with Pt-He and He-He interactions described by Lennard-Jones potential $U_{ij} = A/r_{ij}^{12} - B/r_{ij}^{6}$ ($A$=15.7 eV·Å$^{12}$, $B$=0.989 eV·Å$^6$ for Pt-He and $A$=69.4 eV·Å$^{12}$, $B$=0.494 eV·Å$^6$ for He-He).

The simulation was initialized by randomly putting $n$ Pt atoms and 80 He atoms in a cubic box with a side length of 4 nm (corresponding to a He pressure of about 50 atm at 300 K). Periodic boundary condition was applied, and the temperature of He was controlled at $T$ by replacing all the atomic velocities $\vec{v}_i^{\,old}$ with $\vec{v}_i^{\,new}$ in a time interval of 4 fs [18]. Here,

$$\vec{v}_i^{new} = (1-\theta)^{1/2}\vec{v}_i^{old} + \theta^{1/2}\vec{v}^T, \tag{5}$$

where $\vec{v}^T$ is a random velocity vector chosen from Maxwellian distribution at $T$. The controlling parameter $\theta=0.1$, which could better stabilize the temperature, has been verified in our previous work. The temperature of Pt atoms was set as 2000 K at the beginning of the simulation, and then gradually decreased to $T$ by the effect of He buffer gas and condensed into a cluster. For $T=800\sim1000$ K, the simulation lasted for 100 ns. To monitor the evolution of Pt cluster isomer, the structure was sampled every 5 ps and immediately cooled to 0 K. In once MD simulation, all the cooled samples compose a canonical ensemble of Pt cluster isomers. By counting sample numbers, we got the formation probability of every isomer, and average formation probability at a given temperature $T$ can be derived by repeated simulations. It is worth noting that this technique precludes equilibrating clusters of different sizes, and the following simulation technique is independent of interaction potential and the result was suitable for common cluster growth.

For the smallest magic number $n=13$, the structures and potential energy of 15 isomers with lowest potential energy are shown in Fig. 1(a). The MacKay icosahedron, i.e. the 1st one in Fig. 1(a), was found to have the lowest potential energy and the largest formation probability at $T=800\sim1000$ K. But for other isomers, the one with lower potential energy does not necessarily have higher formation probability [Fig. 1(b)~(d)]. Such situation was also observed for $n=14$, whose 15 ones with lowest potential energy are shown in Fig. 2(a). It can be seen that the 3rd isomer, but not the 1st one, has the largest formation probability [Fig. 2(b)~(d)]. Similar result can be also seen in previous work [4] in which presented a same formation probability spectrum as ours. In fact, for $n=10\sim600$ it was generally found that the formation probability is not closely related to the potential energy. Therefore, the most probable isomer cannot be determined by searching the isomer with the lowest potential energy.

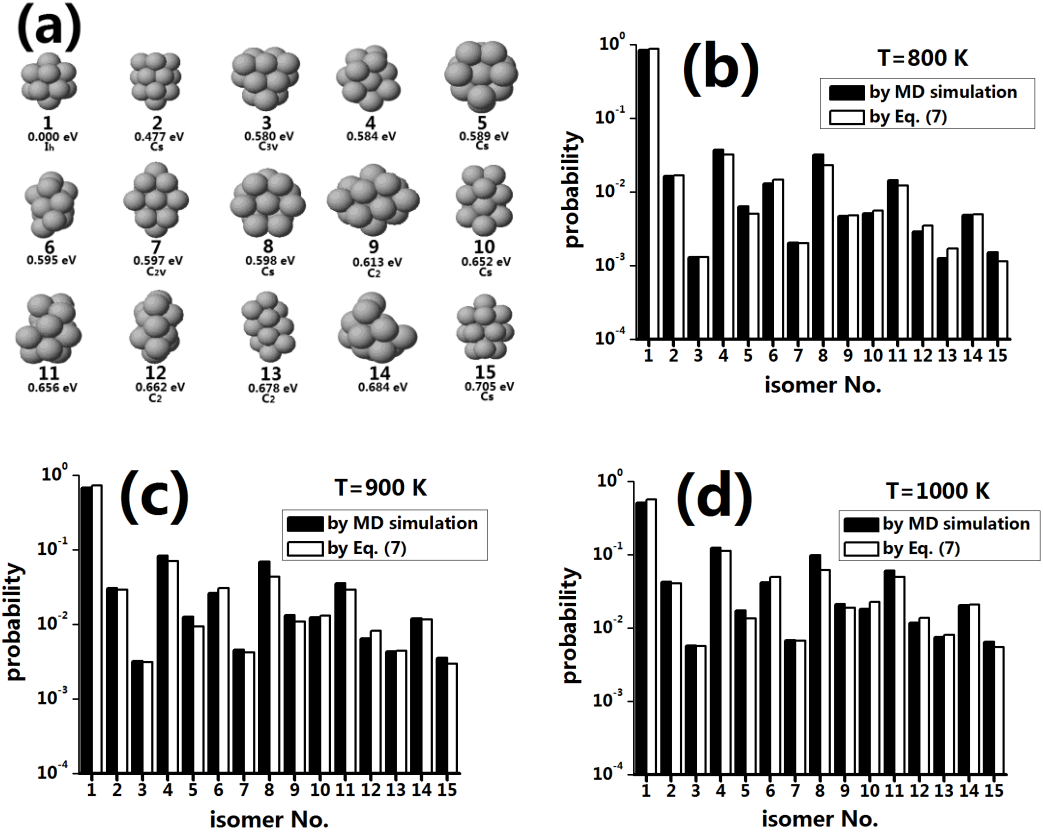

Fig. 1 (a) the 15 isomers of *n*=13 with lowest potential energy. Corresponding potential energy is shown for every isomer, and point group symbols are marked for symmetrical ones. (b), (c) and (d) relative formation probability of the 15 isomers in MD simulations (black column) and calculated by Eq. (7) (white column) at *T*=800, 900 and 1000 K, respectively.

To confirm that the system is in the equilibrium, a technique was applied as follows. MD simulation was performed at given temperature *T*, starting from a selected isomer instead of Pt atomic gas. In the following evolution, frequent isomerization happened and the formation probability of every isomer was derived by counting the samples. If the isomer formation probability spectrum gets close to that in previous MD, it can be judged that stationary equilibrium is achieved in the simulation time. According to the result, for the cases of $n \leq 50$ the spectrum produced from any isomer is similar to the previous one in MD. For example, the formation probabilities of 1st, 4th and 8th isomer of *n*=13 (which are of highest probability) [Fig. 1(a)] in the reproduced spectrum are less different than 2% of the previous one, and such small difference can be eliminated by taking the average of repeated simulations. The same situation was generally found for $n \leq 50$. However, for *n*>50 in once simulation the probability spectrum cannot be in good agreement with the one in previous MD. But by abundant simulations the average formation probabilities produced from any isomer are still close to the previous one, which means for large $n$ a time longer than the simulation duration is needed for the system to achieve stationary equilibrium.

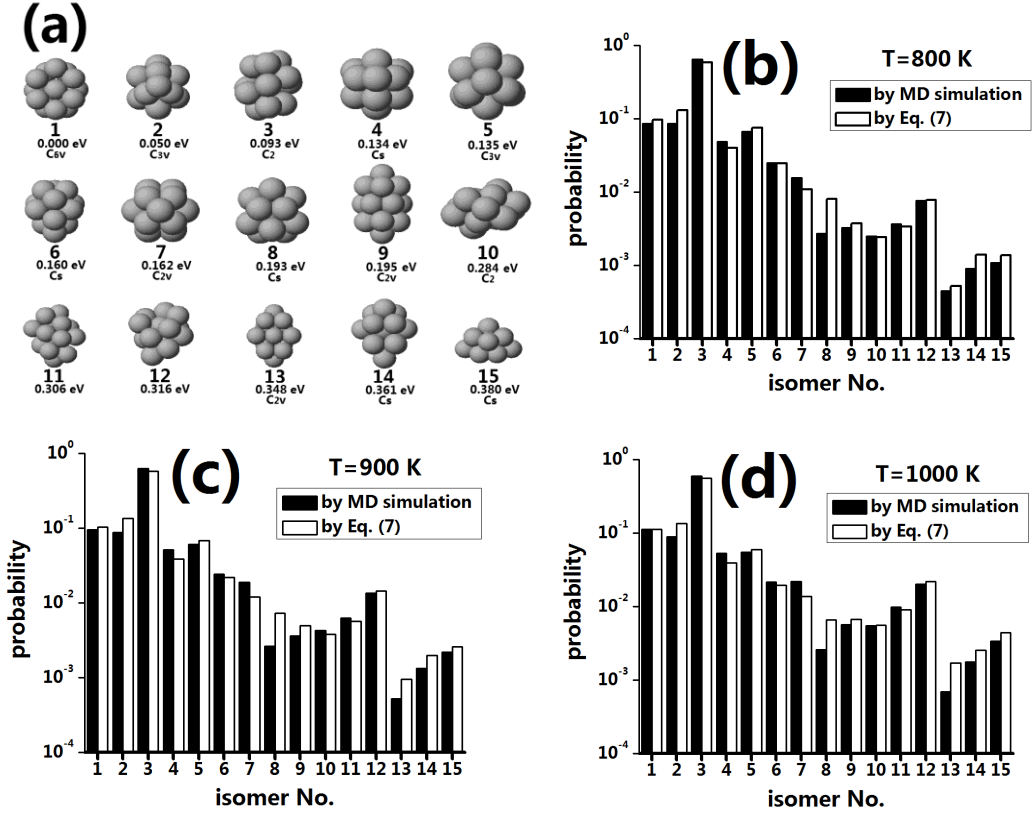

Fig. 2 (a) the 15 isomers of $n=14$ with lowest potential energy. Corresponding potential energy is shown for every isomer, and point group symbols are marked for the symmetrical ones. (b), (c) and (d) the relative formation probability of the 15 isomers in MD simulations (black column) and calculated by Eq. (7) (white column) at $T=800$, 900 and 1000 K, respectively.

In equilibrium, the detailed balance principle Eq. (3) was investigated. Starting from an selected isomer $a$, MD was performed until isomerization happened. By thousands times of simulation, the average rate of isomerization from $a$ to another isomer $b$ was derived. Starting from the 1$^{st}$ isomer of $n=13$, the reactions 1→8 and 1→11 were found, and then from the 8$^{th}$ and 11$^{th}$ isomer the reactions 8→1, 8→11, 8→4, 11→1, 11→8, and 4→8 were found. At $T=900$ K, for the isomerization between the 1$^{st}$ and 11$^{th}$ isomer we have $k_{1st\ 1\to 11}=1.875\times 10^{-5}$ s$^{-1}$ and $k_{1st\ 11\to 1}=3.717\times 10^{-4}$ s$^{-1}$ ($k_{1st\ 1\to 11}/k_{1st\ 11\to 1}$=0.050), which is close to $N_{11}/N_1$=0.052 in MD simulation. Another example is $k_{1st\ 8\to 4}$ =3.175×10$^{-4}$ s$^{-1}$ and $k_{1st\ 4\to 8}$ =3.075×10$^{-4}$ s$^{-1}$, i.e. $k_{1st\ 8\to 4}/k_{1st\ 4\to 8}$=1.033, and in MD we have $N_4/N_8$=1.195. Generally, the detailed balance is satisfied at the simulations at $T$=800~1000 K.

**III. Free energy**

In isothermal-isobaric (or isothermal-isovolumic) ensemble, the isomer number $N$ satisfies Eq. (1) (or Eq. (2)). At temperature $T$, the difference of Gibbs free energy between isomer $a$ and $b$ reads

$$G_b - G_a = F_b - F_a + P_b V_b - P_a V_a = F_b - F_a + kT_b - kT_a = F_b - F_a, \quad (6)$$

where the isomers are treated as ideal gas, and $N_b/N_a$ in isothermal-isobaric or isothermal-isovolumic ensemble is the same. By Eq. (2) and $F = -kT \ln Q$, where $Q$ is the partition function of one molecule, the ratio reads

$$N_b / N_a = e^{-(F_b - F_a)/kT} = Q_b / Q_a. \quad (7)$$

In the following discussion we concern the classical partition function $Q$ to compare with classical MD simulation.

At low $T$, using rigid-rotor and harmonic-oscillator approximation $Q$ can be decomposed as

$$Q = Q_T Q_R Q_V e^{-U_g/kT}, \quad (8)$$

where $U_g$ the potential energy of the isomer and $Q_T$, $Q_R$ and $Q_V$ the translational, rotational and vibrational partition function, respectively. Here,

$$Q_T = \frac{(2\pi M k T)^{3/2} V}{h^3}, \quad (9)$$

where $M$ the molecular mass and $V$ the volume of simulation box, and

$$Q_R = \frac{\pi^3}{h^3 \delta} \sqrt{(8kT)^3 \pi I_x I_y I_z}, \quad (10)$$

where $I_x$, $I_y$ and $I_z$ the molecular principal moment of inertia and $\delta$ the rotational symmetry number. The quantum mechanical expression for the vibrational partition function reads

$$Q_V = \prod_{i=1}^{3n-6} \frac{e^{-h\nu_i/2kT}}{1-e^{-h\nu_i/kT}}, \tag{11}$$

where $\nu_i$ the canonical vibrational frequency of mode $i$. In the classical limit, it becomes

$$Q_V = \prod_{i=1}^{3n-6} \frac{kT}{h\nu_i}. \tag{12}$$

At high $T$, the partition function $Q$ was calculated numerically. For the atoms located at $\vec{r}_1 \sim \vec{r}_n$ with mass $m_1 \sim m_n$ and momentum $\vec{p}_1 \sim \vec{p}_n$, the total energy reads

$$E = \sum_{i=1}^{n} \frac{\vec{p}_i^{\,2}}{2m_i} + U(\vec{r}_1, \vec{r}_2 ... \vec{r}_n) \tag{13}$$

and the classical partition function

$$\begin{aligned} Q &= \frac{1}{h^{3n}\delta} \int e^{-E/kT} d\vec{r}_1 d\vec{r}_2 ... d\vec{r}_n d\vec{p}_1 d\vec{p}_2 ... d\vec{p}_n \\ &= \frac{1}{\delta}\left[\prod_{i=1}^{n}(\frac{\sqrt{2\pi m_i kT}}{h})^3\right]\int e^{-U/kT} d\vec{r}_1 d\vec{r}_2 ... d\vec{r}_n \end{aligned}. \tag{14}$$

The translational motion is separated using a new set of coordinate $\vec{r}_1^{\,'} = \vec{r}_1$, $\vec{r}_2^{\,'} = \vec{r}_2 - \vec{r}_1$, $\vec{r}_3^{\,'} = \vec{r}_3 - \vec{r}_1$ ... $\vec{r}_n^{\,'} = \vec{r}_n - \vec{r}_1$. By $U(\vec{r}_1, \vec{r}_2, \vec{r}_3 ... \vec{r}_n) = U(0, \vec{r}_2^{\,'}, \vec{r}_3^{\,'} ... \vec{r}_n^{\,'})$, the third factor on the right-hand side of Eq. (14) reads

$$\begin{aligned} &\int e^{-U(\vec{r}_1, \vec{r}_2, \vec{r}_3 ... \vec{r}_n)/kT} d\vec{r}_1 d\vec{r}_2 ... d\vec{r}_n \\ &= \int e^{-U(0, \vec{r}_2^{\,'}, \vec{r}_3^{\,'} ... \vec{r}_n^{\,'})/kT} \left|\frac{\partial(\vec{r}_1, \vec{r}_2 ... \vec{r}_n)}{\partial(\vec{r}_1^{\,'}, \vec{r}_2^{\,'} ... \vec{r}_n^{\,'})}\right| d\vec{r}_1^{\,'} d\vec{r}_2^{\,'} ... d\vec{r}_n^{\,'}, \\ &= V\int e^{-U(0, \vec{r}_2^{\,'}, \vec{r}_3^{\,'} ... \vec{r}_n^{\,'})/kT} d\vec{r}_2^{\,'} ... d\vec{r}_n^{\,'} \end{aligned} \tag{15}$$

in which the Jacobian $\frac{\partial(\vec{r}_1, \vec{r}_2 ... \vec{r}_n)}{\partial(\vec{r}_1^{\,'}, \vec{r}_2^{\,'} ... \vec{r}_n^{\,'})} = 1$. Then, the rotational motion is further separated by another coordinate transformation. Starting from $\vec{r}_2^{\,*} = (0,0,r)$, $\vec{r}_3^{\,*} = (\rho, 0, s)$ and arbitrary $\vec{r}_4^{\,*} \sim \vec{r}_n^{\,*}$, any molecular orientation can be produced by 3-2-3 Euler rotation, i.e. rotate $\vec{r}_2^* \sim \vec{r}_n^*$ by $\zeta$ about the z-axis, and by $\theta$ about the y-axis, and then by $\varphi$ about the z-axis. The coordinates $\vec{r}_i^{'} = R\vec{r}_i^*$ generated by the rotation are presented by the rotation matrix

$$R = \begin{pmatrix} \cos\varphi\cos\theta\cos\zeta - \sin\varphi\sin\zeta & -\cos\varphi\cos\theta\sin\zeta - \sin\varphi\cos\zeta & \cos\varphi\sin\theta \\ \sin\varphi\cos\theta\cos\zeta + \cos\varphi\sin\zeta & -\sin\varphi\cos\theta\sin\zeta + \cos\varphi\cos\zeta & \sin\varphi\sin\theta \\ -\sin\theta\cos\zeta & \sin\theta\sin\zeta & \cos\theta \end{pmatrix}. \quad (16)$$

Then, by

$$\vec{r}_2^{'} = R\vec{r}_2^* = R\begin{pmatrix} 0 \\ 0 \\ r \end{pmatrix} = \begin{pmatrix} r\cos\varphi\sin\theta \\ r\sin\varphi\sin\theta \\ r\cos\theta \end{pmatrix} \quad (17)$$

and

$$\vec{r}_3^{'} = R\vec{r}_3^* = R\begin{pmatrix} \rho \\ 0 \\ s \end{pmatrix} = \begin{pmatrix} \rho(\cos\varphi\cos\theta\cos\zeta - \sin\varphi\sin\zeta) + s\cos\varphi\sin\theta \\ \rho(\sin\varphi\cos\theta\cos\zeta + \cos\varphi\sin\zeta) + s\sin\varphi\sin\theta \\ -\rho\sin\theta\cos\zeta + s\cos\theta \end{pmatrix}, \quad (18)$$

the integral element in Eq. (15) reads

$$\begin{aligned}
& d\vec{r}_2^{'} d\vec{r}_3^{'} ... d\vec{r}_n^{'} \\
&= \left| \frac{\partial(\vec{r}_2^{'}, \vec{r}_3^{'}...\vec{r}_n^{'})}{\partial(r,\theta,\varphi,\rho,s,\zeta,\vec{r}_4^*,\vec{r}_5^*...\vec{r}_n^*)} \right| drd\theta d\varphi d\rho ds d\zeta d\vec{r}_4^* d\vec{r}_5^* ...d\vec{r}_n^* \\
&= \left| \frac{\partial \vec{r}_2^{'}}{\partial(r,\theta,\varphi)} \right| drd\theta d\varphi \cdot \left| \frac{\partial \vec{r}_3^{'}}{\partial(\rho,s,\zeta)} \right| d\rho ds d\zeta \cdot \left| \frac{\partial \vec{r}_4^{'}}{\partial \vec{r}_4^*} \right| \left| \frac{\partial \vec{r}_5^{'}}{\partial \vec{r}_5^*} \right| ... \left| \frac{\partial \vec{r}_n^{'}}{\partial \vec{r}_n^*} \right| d\vec{r}_4^* d\vec{r}_5^* ...d\vec{r}_n^* \\
&= r^2 \sin\theta drd\theta d\varphi \cdot \rho d\rho ds d\zeta \cdot d\vec{r}_4^* d\vec{r}_5^* ...d\vec{r}_n^*
\end{aligned} \quad (19)$$

Then, by rotational invariance the final factor in Eq. (15) becomes

$$\begin{aligned}
& \int e^{-U(0,\vec{r}_2^{'},\vec{r}_3^{'}...\vec{r}_n^{'})/kT} d\vec{r}_2^{'}...d\vec{r}_n^{'} \\
&= \int e^{-U(0,\vec{r}_2^*,\vec{r}_3^*...\vec{r}_n^*)/kT} r^2 \sin\theta drd\theta d\varphi \cdot \rho d\rho ds d\zeta \cdot d\vec{r}_4^* d\vec{r}_5^* ...d\vec{r}_n^* . \\
&= 8\pi^2 \int r^2 \rho e^{-U(0,\vec{r}_2^*,\vec{r}_3^*...\vec{r}_n^*)/kT} drd\rho ds d\vec{r}_4^* d\vec{r}_5^* ...d\vec{r}_n^*
\end{aligned} \quad (20)$$

Combining Eq. (14), (15) and (20) we have

$$Q = \frac{8\pi^2 V}{\delta} \left( \prod_{i=1}^{n} (\frac{\sqrt{2\pi m_i kT}}{h})^3 \right) \left( \int r^2 \rho e^{-U(0,\vec{r}_2^*,\vec{r}_3^*...\vec{r}_n^*)/kT} drd\rho ds d\vec{r}_4^* d\vec{r}_5^* ...d\vec{r}_n^* \right) \quad (21)$$

and

$$\frac{Q(T_2)}{Q(T_1)} = \frac{\int r^2 \rho e^{-U/kT_2} dr d\rho ds d\bar{r}_4^* d\bar{r}_5^* ... d\bar{r}_n^*}{\int r^2 \rho e^{-U/kT_1} dr d\rho ds d\bar{r}_4^* d\bar{r}_5^* ... d\bar{r}_n^*}$$

$$= \frac{\int r^2 \rho e^{\frac{U}{k}(\frac{1}{T_1}-\frac{1}{T_2})} e^{-U/kT_1} dr d\rho ds d\bar{r}_4^* d\bar{r}_5^* ... d\bar{r}_n^*}{\int r^2 \rho e^{-U/kT_1} dr d\rho ds d\bar{r}_4^* d\bar{r}_5^* ... d\bar{r}_n^*}, \qquad (22)$$

whose right-hand side can be treated as the average value of $e^{\frac{U}{k}(\frac{1}{T_1}-\frac{1}{T_2})}$ in the canonical ensemble at $T_1$.

Based on the above, a technique was developed to calculate $Q$ at every $T$. At $T=100$ K, $Q$ was calculated by Eq. (8), (9), (10) and (12). Then, Eq. (22) was employed to precisely calculate $Q$ step-by-step from low to high $T$. By Metropolis Monte Carlo method, the calculation temperature $T_1$ was increased to 100, 150, 200 ... 950 K while keeping $T_2=T_1+50$ K. For given $n$, $Q$ of every isomer was calculated and corresponding formation probability was evaluated by Eq. (7). Note, for isomers with chirality, $Q$ was taken as the sum of partition function of two enantiomers.

Fig. 1(b)~(d) and Fig. 2(b)~(d) present the isomer formation probability by Eq. (7) and MD simulation for $n=13$ and 14 at $T=800$, 900 and 1000 K, showing a good agreement between the theoretical and MD value. For $n \leq 50$, the theoretical formation probability was found always in accordance with MD. However, for $n>50$ the isomer formation probability produced in once MD simulation is less consistent with the theoretical value. As an example, for $n=55$ we focus on the sampling of 3 isomers with lowest potential energy, the MacKay icosahedron, and 4 isomers with highest formation probability, which are denoted in sequence as 1~8 [Fig. 3(a)]. For once MD simulation at $T=800$ K, the 4[th] isomer was hardly found [Fig. 3(b)], and for another simulation at $T=900$ K, the 2[nd] and 3[rd] isomer was hardly found [Fig. 3(c)]. Such result indicates that for $n>50$ stationary equilibrium of the system could not be well achieved in the simulation duration (100 ns).

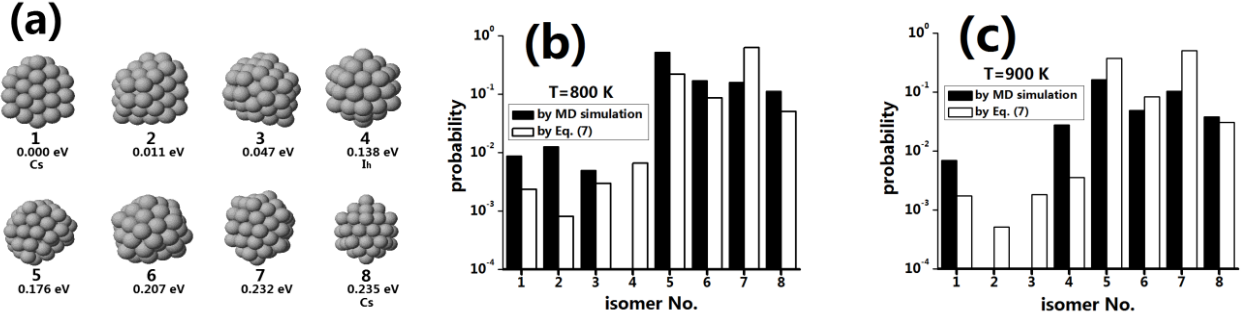

Fig. 3 (a) for *n*=55, the structures of 3 isomers with lowest potential energy, the MacKay icosahedron and 4 isomers with highest formation probability, which are denoted in sequence as 1~8. The potential energy and point group symbols are marked. (b) and (c) the relative formation probability of these isomers in once MD simulation (black column) and calculated by Eq. (7) (white column) at *T*=800 and 900 K, respectively.

## IV. Unimolecular isomerization rate

Although isomer formation probability can be predicted by theoretical calculation of partition function $Q$, the calculation quantity is too high to be applied on *ab initio* simulations since full information of potential energy surface is needed. And by isomerization rate $k_{1st}$, Eq. (3) may be a more convenient way. In this work, a statistical mechanical model was built to evaluate the unimolecular isomerization rate. In nanosystems, an element process may involve transfer of some "key atoms" in a potential valley crossing over $E_0$. In most cases the atomic kinetic energy (~$kT$) at the valley bottom is significantly smaller than $E_0$, and the atom vibrates many times within the valley before crossing over the barrier. In classical level, the Boltzmann distribution of atomic kinetic energy $\varepsilon$ reads

$$f(\varepsilon) = \frac{\varepsilon^{1/2} e^{-\varepsilon/k_B T} d\varepsilon}{\int_0^{+\infty} \varepsilon^{1/2} e^{-\varepsilon/k_B T} d\varepsilon} = \frac{\varepsilon^{1/2} e^{-\varepsilon/k_B T} d\varepsilon}{\sqrt{\pi}(k_B T)^{3/2}/2}, \quad (23)$$

and the probability of $\varepsilon$ larger than $E_0$ is

$$P = \int_{E_0}^{+\infty} f(\varepsilon) d\varepsilon = \frac{\int_{E_0}^{+\infty} \varepsilon^{1/2} e^{-\varepsilon/k_B T} d\varepsilon}{\sqrt{\pi}(k_B T)^{3/2}/2}. \quad (24)$$

With an attempt frequency $k_0$, the atomic transfer rate over the barrier reads

$$k_{1st} = k_0 \frac{\int_{E_0}^{+\infty} \varepsilon^{1/2} e^{-\varepsilon/k_B T} d\varepsilon}{\sqrt{\pi}(k_B T)^{3/2}/2}. \quad (25)$$

Here, $k_0$ can be evaluated by the potential energy $U=U(s)$ along the reaction path,

where $ds^2 = \sum_{i=1}^{n} m_i d\vec{r}_i^2$ is the reaction coordinate. The Lagrangian along the reaction path is

$$L = \frac{1}{2}\left(\frac{ds}{dt}\right)^2 - U, \quad (26)$$

and corresponding Lagrange's equation approximately reads

$$\frac{d^2s}{dt^2} + k_0^2 s = 0, \quad (27)$$

where $k_0 = \left.\frac{d^2U}{ds^2}\right|_{s=0}$ is just the attempt frequency.

To verify the model, its result was compared to TST with rigid-rotor and harmonic-oscillator approximation. In TST, the 1st rate constant reads

$$k_{1st\ a \to b} = \frac{kT}{h}\left(\frac{Q_{TS}}{Q_a}\right)e^{-E_0/kT}, \quad (28)$$

where $E_0$ is the static barrier, $Q_{TS}$ and $Q_a$ are partition functions of transition state and the reactant $a$, respectively. By rigid-rotor and harmonic-oscillator approximation (Eq. (8), (10) and (12)), Eq. (28) becomes

$$k_{1st\ a \to b} = \frac{\delta_a \sqrt{I_{TS,x} I_{TS,y} I_{TS,z}}}{\delta_{TS} \sqrt{I_{a,x} I_{a,y} I_{a,z}}} \left(\frac{\prod_{i=1}^{3n-6} v_{a,i}}{\prod_{i=1}^{3n-7} v_{TS,i}}\right) e^{-E_0/kT}, \quad (29)$$

where $I_{TS}$, $\delta_{TS}$ and $v_{TS}$ are the principal moment of inertia, rotational symmetry number and canonical frequencies of transition state, respectively.

Our model and TST with rigid-rotor and harmonic-oscillator approximation was applied on unimolecular isomerization rate of Pt clusters. For TST, molecular canonical frequency and moment of inertia were calculated by the potential Eq. (4). By nudged elastic band method [19-21], the potential energy $U=U(s)$ along the reaction path was obtained to apply our model (Eq. (25)). To obtain enough data to verify TST and our model, MD simulations for typical isomerization progresses were repeatedly performed at $T=700 \sim 1000$ K by thousands of times, and the derived $k_{1st}$ were averaged at every temperature. Then, we changed the side length of simulation box in 2~15 nm and performed the same MD simulations, finding that at every temperature the isomerization rate are independent of the side length of simulation box, which means the molecular concentration of He buffer gas is high enough and the isomerization progresses are all 1st-order reactions.

For 5 typical isomerization progresses of $n$=13, $E_0$, $k_0$ and reaction path degeneracy are shown in Table 1. Here, the atomic migration barrier $E_0$ in small Pt clusters are smaller than the atomic migration barrier in surface islands of Pt solid [15]. Note, 60 equivalent reaction paths were found for 1→8 and 1→11 since the 1st isomer has Ih symmetry. Fig. 4(a)~(c) shows the result of MD, TST and our model for these reactions. For 1→8, 11→1 and 8→4, both TST and our model are in accordance with the MD data, while having some deviation for 8→1. For 1→11, our model is better than TST. Generally, our model is successful in predicting isomerization rate of Pt clusters with $n$=13.

|  | 1→8 | 8→1 | 1→11 | 11→1 | 8→4 |
| --- | --- | --- | --- | --- | --- |
| $E_0$ (eV) | 0.819 | 0.221 | 0.794 | 0.138 | 0.121 |
| $k_0$ (s$^{-1}$) | $3.18 \times 10^{12}$ | $2.38 \times 10^{12}$ | $4.42 \times 10^{12}$ | $9.25 \times 10^{11}$ | $1.02 \times 10^{12}$ |
| reaction path degeneracy | 60 | 1 | 60 | 1 | 1 |

Table 1 $E_0$, $k_0$ and reaction path degeneracy of some Pt cluster isomerization reactions of $n$=13. The isomer numbers correspond to Fig. 1(a).

For further verification, MD simulation was performed for Mackay icosahedron of Pt cluster with $n$=55, and the rate data of most probable progress (see the sketch in Fig. 4(c)) was employed. By the reaction path calculation, we got $E_0$=0.674 eV and $k_0$=1.39$\times 10^{12}$ s$^{-1}$, and the reaction path degeneracy of Mackay icosahedron with Ih symmetry is 60. The corresponding $k_{1st}$ is smaller than $10^9$ s$^{-1}$ at $T$=500 K, which is in accordance of previous simulation [15]. For $T$=700~1000 K, $k_{1st}$ derived by MD, TST and our model were plotted in Fig. 4(c), showing the validity of both TST and our model.

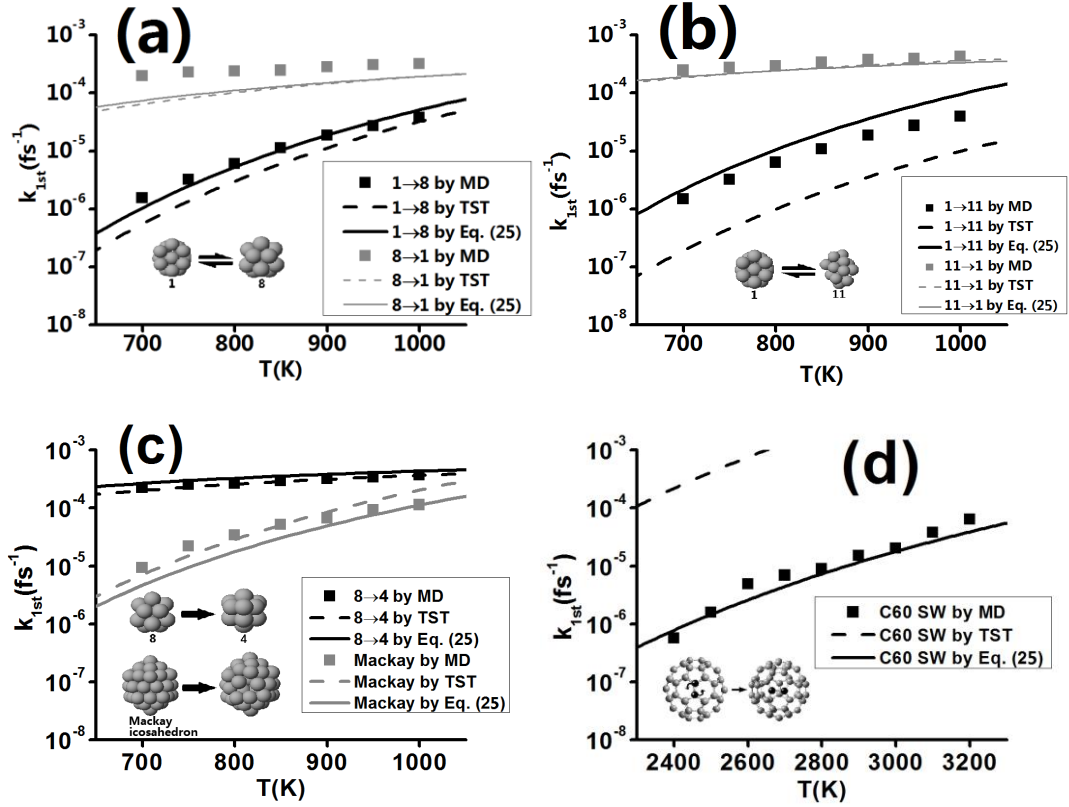

Fig. 4 For Pt clusters: (a) $k_{1st}$ for $1 \rightleftharpoons 8$ of $n=13$; (b) $k_{1st}$ for $1 \rightleftharpoons 11$ of $n=13$; (c) $k_{1st}$ for $8 \rightarrow 4$ of $n=13$ and isomeriation of Mackay icosahedron of $n=55$. For $C_{60}$ fullerene: (d) for Stone-Wales transformation. In above figures, squares, dashed lines and solid lines are for by MD, TST and our model, respectively.

Finally, our model was applied on the Stone-Wales transformation of $C_{60}$ fullerene (see the sketch in Fig. 4(d)). MD simulation was performed by the same technique used above with C-C interaction described by the Brenner potential [22]. By reaction path calculation, we got $E_0=3.268$ eV and $k_0=1.05 \times 10^{13}$ s$^{-1}$. The reaction path degeneracy is 120 because the 60 atoms in $C_{60}$ molecule are equivalent and each atom has two vibration directions for the reaction. For $T=2400 \sim 3200$ K, $k_{1st}$ derived by MD, TST and our model were plotted in Fig. 4(e), showing that our model fits the MD data well while TST fails to predict the rate constant.

## V. Summary

In summary, based on free energy calculation, theoretically predicted isomer formation probability of Pt clusters is in good agreement with MD simulations for cluster gas-phase growing. And the detailed balance between isomers was verified by MD. For clusters of $n \leq 50$, stationary equilibrium is achieved in 100 ns in the canonical ensemble, while longer time is needed for $n>50$. Then, a statistical mechanical model was built to evaluate isomerization rate and simplify the prediction of isomer formation probability. By MD data, its accuracy was validated. This model is simpler than TST and can be easily applied on *ab initio* calculations to predict the lifetime of nanostructures.

***

**Acknowledgements**

This work was supported by the National Natural Science Foundation of China under Grant No. 11304239, and the Fundamental Research Funds for the Central Universities.